\documentstyle[prl,aps,epsf]{revtex}
\begin{document}
\twocolumn[\hsize\textwidth\columnwidth\hsize\csname %
@twocolumnfalse\endcsname

\draft
\title{Aharonov-Anandan Effect Induced by Spin-Orbit Interaction 
and  Charge-Density-Waves in Mesoscopic Rings}
\author{Ya-Sha Yi and A. R. Bishop}
\address{Theoretical Division and Center for Nonlinear Studies, 
Los Alamos National Laboratory, Los Alamos, NM 87545}
\date{November 26, 1997}
\maketitle
\begin{abstract}
We study the spin-dependent geometric phase effect in   
mesoscopic rings of charge-density-wave(CDW) materials. When 
electron spin is explicitly taken into account,  
we show that the spin-dependent 
Aharonov-Casher phase can have a pronounced frustration effects on such 
CDW materials with appropriate electron filling. We show that this
frustration has observable consequences 
for transport experiment. We identify a 
phase transition from a Peierls insulator to metal, which 
is induced by spin-dependent phase interference effects.  
Mesoscopic CDW materials and  spin-dependent 
geometric phase effects, and  their interplay,  
are becoming attractive opportunities for exploitation with the 
rapid development of modern fabrication technology. 

\end{abstract}
\phantom{.}

]

\narrowtext
\pagebreak

Since Berry identified the importance of a "geometric phase" 
in adiabatic cyclic 
evolution\cite{1}, there have been numerous theoretical and experimental 
studies of this holonomy phenomenon termed the Berry phase\cite{2}. 
A fundamental generalization of the Berry phase was given by Aharonov and 
Anandan(AA)\cite{3}. They removed the adiabatic condition and demonstrated 
the existence of the geometric phase in generic cyclic evolution. 
As is well known, the Aharonov-Bohm(AB) effect has led to a number of 
remarkable interference phenomena in mesoscopic systems, especially 
in rings\cite{4}. Based on the discovery of the geometric phase, it has been 
predicted that analogous interference effects can be induced by the geometric 
phases which originate from the interplay between an electron's {\it spin} and 
{\it orbital} degrees of freedom. Such an interplay can be maintained by an 
external 
electric field, which leads to a spin-orbit(SO) interaction.  

Loss et al. first studied a textured ring  in an inhomogeneous magnetic 
field\cite{5}. They found that the inhomogeneity of the field can results in a Berry 
phase, which can result in persistent currents. The effects of this Berry phase 
on conductivity were then discussed\cite{6}. On the other hand, the 
Aharonov-Casher(AC) effect\cite{7} in mesoscopic systems has also attracted 
much attention, since it specifically includes the spin degree of freedom. 
Meir et al. showed that SO interaction in one 
dimensional rings results in an effective magnetic flux\cite{8}. Mathur and 
Stone then pointed out that observable phenomena induced by SO interactions  
are essentially the manifestation of the AC effect  
and proposed an observation of the AC oscillation of 
the conductance on semiconductor samples\cite{9}. Balatsky and 
Altshuler\cite{11} and Choi\cite{12} studied the persistent current produced 
by the AC effect. Inspired by these studies of textured rings, the AC effect 
has also been analyzed in connection with the spin geometric phase. 
Aronov and Lyanda-Geller considered the spin evolution in conducting rings, 
and found that SO interaction results in a spin-orbit Berry phase which 
plays an interesting role in the transmission probability\cite{13}. 

The charge-density-wave(CDW) broken-symmetry state induced 
by electron-phonon interaction has also been intensively investigated 
during the last decades. The dynamics of CDWs in 
materials such as 
$NbSe_3$\cite{14}, as well as their collective excitations\cite{15}, 
have received detailed study.  
Recently, it has been found that an external magnetic field has a pronounced 
effect on the CDW ground state\cite{16,17}. 
In  sufficiently small mesoscopic rings, the AB flux induced 
by an external magnetic field can even destroy the CDW ground state\cite{17}:  
for the first time the instability of the CDW ground state with 
respect to the AB effect has been shown. Recently, it is
further emphasized that, for spinless electrons, the AB
effect depends on the parity of the number $N$ of
electrons\cite{17b}. Specifically, when the number of spinless electrons
are even, the electronic polarizability, 
which in the absence of magnetic flux has a well known divergence at 2$k_F$, 
can be compensated by magnetic flux , which then has a similar effect to  
temperature, inducing a transition from Peierls distortion 
to metal. 

In this paper, we focus on the role of the electron {\it spin} in a  
{\it cylindrical electrical}  field, which is the source of  
the SO interaction.  This induces 
an AC phase, as the electrical field is dual to the 
magnetic field. We concentrate on the condition which has a
{\it destruction} effect on the mesoscopic CDW system. We found that, when 
{\it spin} degree of freedom is explicitly taken into account, the 
parity effect is more complicated and 4$n$($n$ is a integer) electrons 
has definite destruction effects, which is quite different from 
the spinless case, we will address the other filling case elsewhere. In
the following, we focus on the 4$n$ electron case, 
since our interest is mainly on the sector in which the spin-dependent
geometric phase has a {\it destruction} effect.   
When the spin
and the spin-dependent geometric phase are explicitly taken into account, 
we show that the Aharonov-Casher phase(comprised of 
the nonadiabatic AA phase and the dynamical phase by SO interaction) induced 
by the cylindrical electric field can have  a pronounced
destruction effect on the 
CDW. We further propose novel  
observable consequences on the 
transport properties of mesoscopic CDW rings. 

\begin{figure}
\vskip -10pt
\centerline{
\epsfxsize=7.0cm \epsfbox{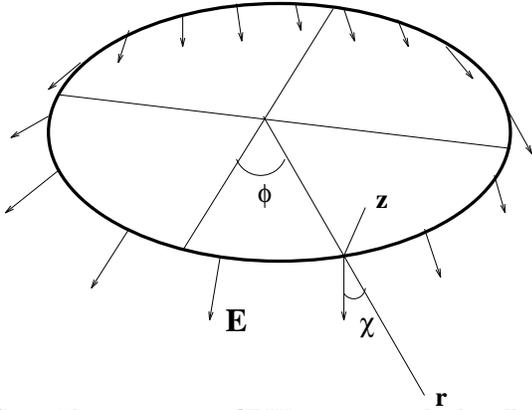}}
\vskip -0pt
\caption{The mesoscopic CDW ring in a cylindrically symmetric field 
with tilt angle $\chi$.}
\end{figure}

In the presence of an electric field ${\bf E}=-\nabla V$, 
the one-particle Hamiltonian for non-interacting electrons 
confined to a mesoscopic ring is : 
\begin{equation}
H=\displaystyle\frac {1}{2m_{e}}{\bf p}^{2}
+eV-\displaystyle\frac{e\hbar}{4m_{e}^{2}c^{2}}\sigma\!\!\!\!\sigma\cdot{\bf E}
\times {\bf p}, 
\end{equation}
Where ${\bf \sigma}$ is the Pauli matrix, $m_{e}$ is the effective mass 
of electrons and ${\bf p}$ represents the momentum of electrons.  
We consider a ring that is effectively one-dimensional(1D) and where the 
electric field which results in the SO interaction is cylindrically 
symmetric( see Fig.1), i.e.,  
${\bf E}=E(\cos\chi {\bf e_{r}}-\sin\chi {\bf e_{z}})$.  
For a ring lying in the $xy$ plane with its center at the 
origin, the Hamiltonian reads  
\begin{equation}
H= \displaystyle\frac{\hbar^{2}}{2m_{e}a^{2}}[-i\displaystyle
\frac{\partial}{\partial \theta}+\alpha(\sin\chi\sigma_{r}
+\cos\chi\sigma_{z})]^{2}-\displaystyle\frac{\alpha^{2}
\hbar^{2}}{2m_{e}a^{2}},
\end{equation}
with $\sigma_{r}=\sigma_{x}\cos\theta +\sigma_{y}\sin\theta $ and 
$\alpha =-\displaystyle\frac{eaE}{4m_{e}c^{2}}$, 
where $a$ is the ring radius, and $\theta$ is the angular coordinate. 

We adopt the geometric phase approach to identify the 
geometric and dynamical phases\cite{3}, which are responsible for 
the effects on the CDW broken-symmetry ground state in mesoscopic rings. 
The eigenstates of the Hamiltonian are of the form 
\begin{equation}
\Psi_{n,\mu}(\theta)=\exp(in\theta)\tilde{\psi}_{n,\mu}(\theta)/\sqrt{2\pi}, 
\end{equation}
in which $\mu=\pm$, $n$ are arbitrary integers, and the spin states are given by
\begin{equation} 
\tilde{\psi}_{n,+}(\theta)=\left[
\begin{array}{c} \cos\displaystyle\frac{\beta}{2}\\
		 e^{{\it i}\theta}\sin\displaystyle\frac{\beta}{2}\\
\end{array} \right ];\;\;\;
\tilde{\psi}_{n,-}(\theta)=\left[
\begin{array}{c} -e^{{\it -i}\theta}\sin\displaystyle\frac{\beta}{2}\\
		 \cos\displaystyle\frac{\beta}{2}\\
\end{array} \right ]
\end{equation}
with $\tan\beta = \displaystyle\frac{2\alpha\sin\chi}{2\alpha\cos\chi - 1}$.
The geometrical phase(AA phase) is given by
\begin{equation}
\int_{0}^{2\pi}i\tilde{\psi}^{(\mu)^{\dagger}}(\theta)
d \tilde{\psi}^{(\mu)}(\theta)=-\mu\pi(1-\cos\beta), 
\end{equation}
and the dynamical phase is 
\begin{equation}
-\int_{0}^{2\pi}i\tilde{\psi}^{(\mu)^{\dagger}}(\theta)
H_{s} \tilde{\psi}^{(\mu)}(\theta) d\theta=
-2\mu\pi\alpha\cos(\beta-\chi). 
\end{equation}
Then the AC phase is  
\begin{equation}
\phi_{AC}^{\mu}=-\mu\pi(1-\cos\beta)-2\mu\pi\alpha\cos(\beta-\chi), 
\end{equation}
which satisfies $\sum_{\mu}\phi_{AC}^{\mu}=0$,  
and the solution of the Hamiltonian in the mesoscopic system is 
\begin{equation}
\varepsilon_{n,\mu}= \displaystyle\frac{\hbar\omega_{0}}{2}
(n- \frac{\phi_{AC}^{\mu}}{2\pi})^{2}
-\frac{\alpha^{2}\hbar^{2}}{2m_{e}a^{2}}, 
\end{equation}
where $\omega_{0}=\hbar/ma^{2}$. 

The AC phase comprises the geometric AA phase and the dynamical 
phase, which is obtained by the spin cyclic evolution of the spin 
freedom of the electron. This AC phase will change 
the wave numbers of the two 
independent spin polarized cyclic electron gases.  
Thus, when the spin degree-of-freedom is explicitly taken into 
account, the accumulated {\it spin}-dependent geometric phase 
will have pronounced effects in a CDW mesoscopic ring, and 
result in an interesting  effect on 
the transport properties of the CDW system, as we discuss 
in the following.

As is well known, electron-phonon interaction treated adiabatically  
in a quasi-one dimensional system leads to a CDW gap at wavevector 
$q=2k_F$, so that 
the originally continuous energy band breaks into two bands:   
valence and  conduction. Since Peierls first 
pointed out that a one-dimensional 
metal coupled to the lattice  
is unstable at low 
temperature, both theoretical and experimental studies have 
concentrated on the static and dynamical characters of 
charge-density-waves, including the frequency- and electric-field 
-dependent conductivity, current oscillation and pinning via  
defects or disorder\cite{14}. These studies 
focused on the conductivity character in direct external electrical 
fields for macroscopic quasi-one dimensional CDW materials, such as 
$K_{0.3}MoO_{3}$, $NbSe_{3}$, etc. As fabrication techniques have 
become more mature, it is now promising that fabrication of mesoscopic CDW 
samples can be realized\cite{18}. 

The logarithmical 
singularity of the dielectric response function at  
wavevector $2k_{F}$ makes the corresponding phonon frequency   
soften drastically(Kohn anormaly). The modulated 
lattice structure also affects the electronic band 
structure. In second-quantized representation,  
with a cylindrical electric field in a CDW mesoscopic ring, 
we can use the  Hamiltonian 
\begin{equation}
H=\sum_{k,\mu}(\varepsilon_{k,\mu}C_{k,\mu}^{\dagger}C_{k,\mu}
+\Delta (C_{k+2k_{F},\mu}^{+}C_{k,\mu}+ H.C.)) . 
\end{equation}
Here the CDW gap is $\Delta=2\gamma u$, with $\gamma$ the 
electron-lattice coupling constant and $u$ the displacement 
after dimerization,  and  $\varepsilon_{k,\mu}$ 
the eigenenergies in an external cylindrical electric 
field(Eq.8). We consider the half-filled case, but  
extension to other band fillings for  many real quasi-one dimensional 
CDW materials is straightforward. After diagonalizing the reduced 
2x2 matrix 

\begin{equation}
\left\| 
\begin{array}{cc}
E_{k,\mu}-\varepsilon_{k,\mu} & \Delta \\
	\Delta & E_{k,\mu}-\varepsilon_{k+2k_{F},\mu}
\end{array}
\right\| =0,  
\end{equation}
we obtain the splitting into valence and conduction bands:  
\begin{equation}
\begin{array}{cl}
E_{k,\mu}^{val}=& \displaystyle\frac{1}{2}(\varepsilon_{k,\mu}
+\varepsilon_{k+2k_{F},\mu})-\frac{1}{2}\sqrt{(\varepsilon_{k,\mu}
-\varepsilon_{k+2k_{F},\mu})^2+4\Delta^{2}} \\
E_{k,\mu}^{con}=& \displaystyle\frac{1}{2}(\varepsilon_{k,\mu}
+\varepsilon_{k+2k_{F},\mu})+\frac{1}{2}\sqrt{(\varepsilon_{k,\mu}
-\varepsilon_{k+2k_{F},\mu})^2+4\Delta^{2}} .\\
\end{array}
\end{equation}
From Eq.(11), we see that when the spin degree-of-freedom 
is taken into account, 
the spin-dependent geometric phase affects the CDW ground 
state. First, we concentrate on the 
effect on the CDW gap. We will turn to the effect on transport 
properties later. As the $2k_{F}$ mode is dominant 
in the electron-phonon interaction, we can obtain the following total 
effective potential:
\begin{equation}
E=\sum_{|k|\leq k_{F}}E_{k,\mu}^{val}+\displaystyle\frac{1}{2}
\omega_{2k_{F}}^{2}u_{2k_{F}}^{2}, 
\end{equation}
where $\Delta=\gamma u_{2k_{F}}$. Adopting a  
standard minimization proceedure , we now find the effect on the 
CDW gap by the spin-dependent geometric phases for the 
electron numbers 4$n$ are: 
\begin{equation}
\prod_{\mu}\displaystyle\frac{|\frac{\phi_{AC}^{\mu}}{2\pi}|+
\sqrt{(\frac{\phi_{AC}^{\mu}}{2\pi})^{2}+(\frac{\Delta}
{\hbar\omega_{0}n_{F}})^{2}}}{|\frac{\phi_{AC}^{\mu}}{2\pi}+
2n_{F}|+\sqrt{(\frac{\phi_{AC}^{\mu}}{2\pi}+2n_{F})^{2}
+(\frac{\Delta}{\hbar\omega_{0}n_{F}})^{2}}}=\exp(-\frac{1}{g}), 
\end{equation}
with $g$ the dimensionless effective electron-lattice interaction, 
$g=\gamma^{2}/\hbar\omega_{0}n_{F}m\omega_{2k_{F}}^{2}$. 

\begin{figure}
\vskip -18pt
\centerline{
\epsfxsize=7cm \epsfbox{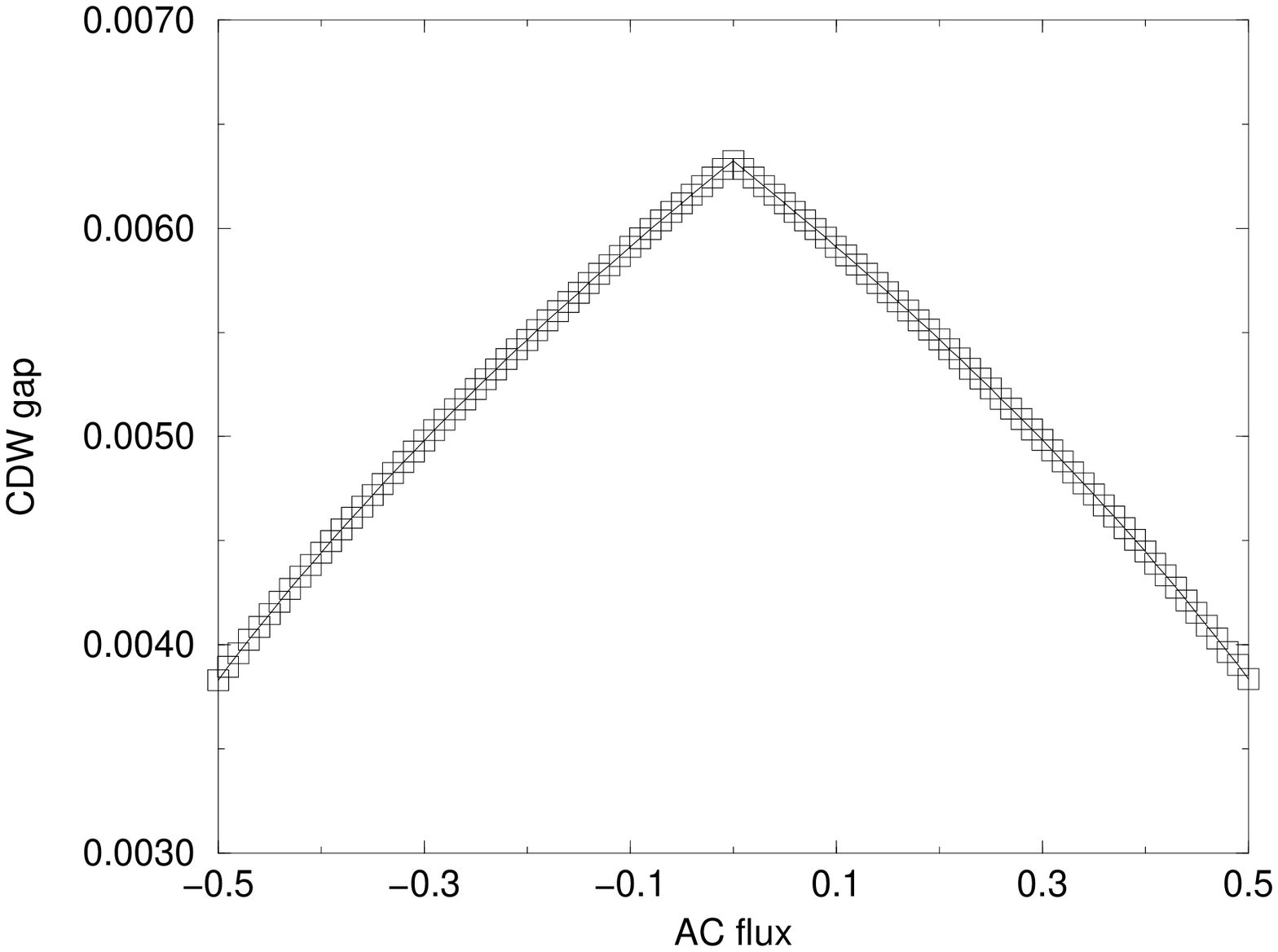}}
\vskip -10pt
\caption{The CDW gap dependence on the spin-dependent geometric 
phase, for the strength of effective e-ph interaction g=0.07.}
\end{figure}

\begin{figure}
\vskip -20pt
\centerline{
\epsfxsize=7cm \epsfbox{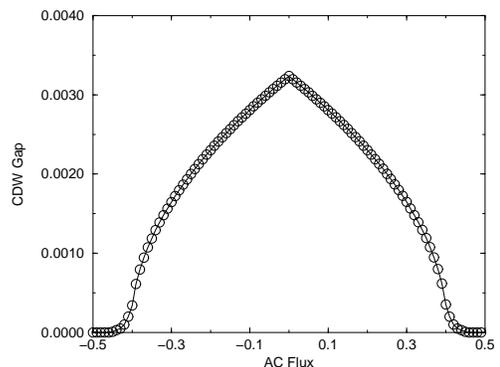}}
\vskip -10pt
\caption{The CDW gap dependence on the spin-dependent geometric 
phase, for the strength of effective e-ph interaction g=0.064.}
\end{figure}


We have explored this AC phase effect on a CDW mesoscopic ring in various 
regions. We found that it is very sensitive to e-ph coupling, 
as illustrated in  
Figs.2,3. The effects of the spin-dependent geometric phases for 
different e-ph coupling constants shows that only when the 
e-ph coupling is weak enough, is a breaking effect on the CDW  
observable:   
the stronger the e-ph coupling, the more stable is the CDW. 
We emphasize that the breaking effect(4$n$ electrons) by the AC phase 
is the result of the two spin-polarized electrons, which is 
produced by the external electric field, and these 
two spin polarized electron gases accumulate a spin-dependent 
geometric phase through the SO interaction. Hence, we 
have identified 
a new mechanism, different from that due to an 
external magnetic field.

We have concluded that when the external electrical field reaches a 
critical strength, the CDW ground state in a mesoscopic ring can be 
destroyed with the appropriate electron fillings(4$n$ when spin is
explicitly 
taken into account). Since the SO interaction is time-reversal-invariant,
we expect 
that the effect of the spin-dependent geometric phase on a CDW mesoscopic 
ring will  be manifested in transport processes. Here, it is 
induced by the cylindrical electric field. This is different from 
previous studies in which the electrical field is parallel to 
the direction of the quasi-one dimension of the CDW materials, and 
hence has no corresponding topological effect on 
the transport.  We adopt the configuration  
with the ring connected to two current leads in opposite 
directions, which is the standard structure for transport 
studies and interference effects in a mesoscopic field(see Fig. 4).  

\begin{figure}
\vskip -12pt
\centerline{
\epsfxsize=7.0cm \epsfbox{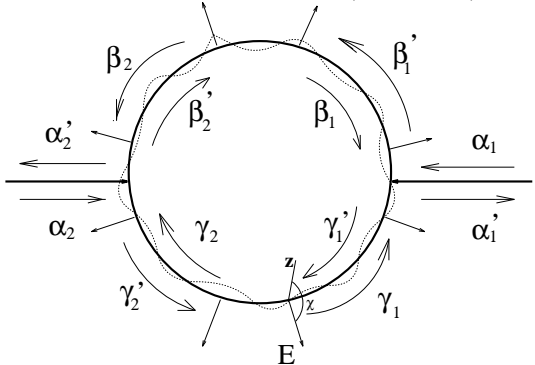}}
\vskip +2pt
\caption{Schematic illustration of the electronic waves propagating 
through a ring connected to current leads. The right junction is located 
at $\theta=0$ and the left junction at $\theta=\pi$ with the upper 
branch lying within (0,$\pi$) and the lower one within ($\pi$,2$\pi$).
The dotted line illustrates the charge-density-wave in the ring before 
the symmetry breaking.}
\end{figure}

Our formulation is standard for that developed in the study of quantum 
oscillations\cite{19}.  
We obtain the transmission 
probability in the presence of SO interactions(a detailed 
derivation will be given elsewhere\cite{20}) as:

\begin{equation}
T= \displaystyle\frac{1}{2}\sum_{\mu}t(\frac{\phi_{AC}^{\mu}}{2\pi}), 
\end{equation}
with 
$$t(\frac{\phi_{AC}^{\mu}}{2\pi})=\displaystyle\frac{4\epsilon^{2}
\sin^{2}\phi_{s}\cos^{2}\phi_{AC}^{\mu}}{[a^{2}+b^{2}\cos2\phi
_{AC}^{\mu}-(1-\epsilon)\cos2\phi_{s}]^{2}+\epsilon^{2}\sin^{2}
2\phi_{s}} .$$  
Here $\phi_{s}$ is the phase of the incident electron on the lead,  
where $a=\pm (\sqrt{1-2\epsilon}-1)/2$ and $b=\pm (\sqrt{1-2\epsilon}+1)/2$
with $0\leq\epsilon\leq 1/2$\cite{19}.

\begin{figure}
\vskip -10pt
\centerline{
\epsfxsize=7cm \epsfbox{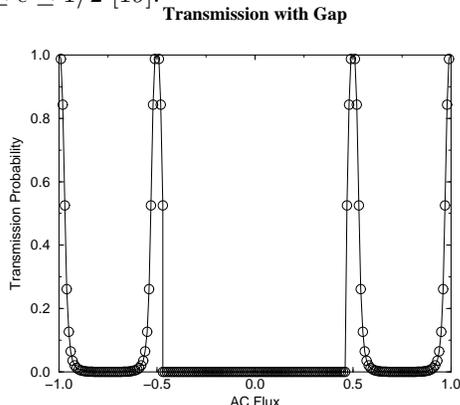}}
\vskip -10pt
\caption{The transmission probability with respect to the 
spin-dependent geometric phase, when the incident electron energy 
is within the CDW gap, and with g=0.064.}
\end{figure}

\begin{figure}
\vskip -10pt
\centerline{
\epsfxsize=7cm \epsfbox{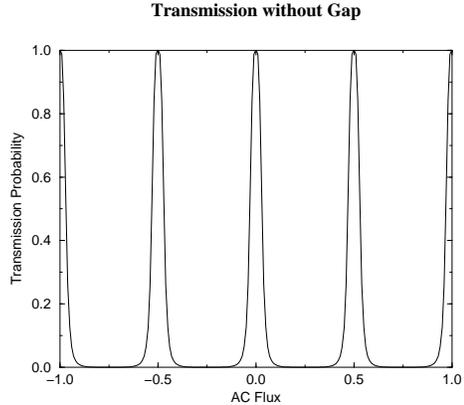}}
\vskip -10pt
\caption{The transmission probability with respect to the 
spin-dependent geometric phase in the absence of a CDW gap.}
\end{figure}

From Fig.5, we see that the CDW breaking effect by the spin-dependent 
geometric phase is clearly manifested in electronic transport.  
When the CDW is absent, the transmission probability 
of the conducting ring should be periodic with respect to the 
AC phase, as illustrated in Fig.6 . When the e-ph coupling is turned 
on and the energy of the incident wave is within the CDW gap, a clear   
effect can be seen with the variation of the AC phase.  
Hence, there exists a critical value of 
$\phi_{AC}^{\mu}$($\pm 0.43$ for the effective e-ph coupling 
$g=0.064$),  where 
the transmission probability of the incident energy has a large jump, 
as illustrated in Fig.5, a signature of the transition 
from Peierls insulator to metal. For experimental convenience, 
we can take 
the direction of the electrical field along the {\bf z} axis(Fig.1), 
i.e., $\chi=\pi/2$. In a mesoscopic ring with radius $a=100\mu m$, 
to make the above effect observable, an electrical field 
with strength $\sim 10^6 V/m$ is necessary, so that the AC phase can 
be the order of unity.   

In summary, we have investigated a spin-dependent geometric phase effect  
in mesoscopic CDW rings. When the electron spin is explicitly 
taken into account in the presence of a cylindrical external 
field and under appropriate filling electrons, the AC(AA) phase
accumulated by the two independent 
spin polarized electron gases can results in frustration of the 
CDW on a mesoscopic scale. We thus propose a new mechanism 
with which to probe mesoscopic CDW materials, and associated  
spin-dependent geometric 
phase consequences. As a novel consequence, we suggest that 
a frustration effect of the spin-dependent geometric phase will be 
observable in transport experiments. 
There are natural extensions of our considerations to 
competing spin-density-wave and CDW situations, as well as the 
other filling electron case,  which we
will 
discuss elsewhere. These opportunities for studying mesoscopic 
systems are becoming increasingly 
attractive with the rapid progress of  
modern fabrication technology.    
 
This work was supported by the U. S. Department of Energy.  
One of us(Y. S. Yi) thanks Prof. Z. B. Su and Dr. T. Z. Qian for 
helpful discussions.

\end{document}